# Platform for enhanced light-graphene interaction length and miniaturizing fiber stereo-devices

Jun-long Kou,[‡] Jin-hui Chen,[‡] Ye Chen, Fei Xu,[*] and Yan-qing Lu[*]

*feixu@nju.edu.cn and yqlu@nju.edu.cn

National Laboratory of Solid State Microstructures and College of Engineering and Applied Sciences, Nanjing University, Nanjing 210093, China

ABSTRACT: Sufficient light-matter interactions are important for waveguide-coupled graphene optoelectronic devices. Using a microfiber-based lab-on-a-rod technique, we present a platform for ultra-long light-graphene interaction and design graphene-integrated helical microfiber (MF) devices. Using this approach, we experimentally demonstrate an in-line stereo polarizer by wrapping an MF on a rod pretreated with a graphene sheet. The device operates as a broadband (450 nm wavelength) polarizer capable of achieving an extinction ratio (ER) as high as ~8 dB/coil in the telecommunication band. Furthermore, we extend this approach to successfully demonstrate a high-Q graphene-based single-polarization resonator, which operates with an ER of ~11 dB with excellent suppression of polarization noise. The fiber-coil resonator shows great potential for sensing applications and gyro-integration. By specializing the rod surface and coil geometry, we believe the preliminary results reported herein could contribute to advancing the research for lab-on-a-rod graphene-MF-integrated devices.

Since the discovery of graphene, the optical properties of two-dimensional Dirac fermions have been studied extensively.[1,2] By utilizing properties such as linear optical absorption, saturable absorption, and the tunability of chemical potential through doping or electrical gating, various broadband applications, such as modulators,[3,4] photon detectors,[5,6] polarizers,[7] and mode-lock lasers,[8,9] have been realized. Most practical applications require graphene to be integrated with existing photonic technology to attain sufficient interaction lengths between the graphene and optical field because the interaction length is limited by the thickness of graphene for normal incident light in conventional, free-space optical systems. This integration can be readily achieved by transferring and laminating graphene on top of waveguides.[10] In fiber-optic systems, side-polishing or flame-drawing techniques are employed to obtain D-shaped fibers or microfibers (MFs),[7,11] which provide an accessible evanescent field. In particular, MFs with a strong evanescent field have attracted increasing attention. Hybrid graphene-microfiber (GMF) devices can be fabricated by covering or wrapping a graphene sheet on a straight, thin MF.[12] However, it is still challenging to handle such a thin MF-graphene structure for sufficient lengths and strengths of interaction. Below, relying on the MF-based lab-on-a-rod technique, we present an alternative approach to integrate GMF devices that is based on wrapping an MF on a graphene-coated rod, as shown in Figure 1. Implementation of the approach is simple and efficient because the process only involves coating a small piece of graphene onto a rod. While maintaining a strong evanescent field, the GMF interaction length can, in theory, be increased arbitrarily with a helical structure of multi-coils. Another unique advantage is the possible formation of resonators with strong coupling between adjacent coils through adjustment of the spring pitches. The approach provides new opportunities for elaboration of stereo GMF-integrated devices. Various all-fiber graphene devices are expected to be developed by employing such a flexible platform; such devices include electrical or optical modulators, photon detectors, and polarization controlling components (PCCs). In this work, we show the platform's application to broadband polarization manipulation.

Polarization behavior has a profound impact on the performance of optical-fiber devices and systems.[7, 13] In order to implement optical-fiber devices in practical applications, control and manipulation of the polarization state of light is highly desirable. All-fiber PCCs are more attractive because manipulating free-space PCCs during collimating, aligning, and (re)focusing is time-consuming and labor-intensive. However, conventional-fiber PCCs have to be carefully kept as straight as possible to prevent additional birefringence. This requires splicing or connecting and careful assembly as well as alignment to achieve the desired optical performance. It is also difficult to coil conventional-fiber PCCs well and form a single-polarization microresonator.

In this study, a stereo GMF-integrated platform and the so-called wrap-on-a-rod technique are used to enable the integration and miniaturization of multifunctional GMF PCCs. The key to the technique lies in the graphene specialization of the rod's surface, which interacts strongly with the evanescent field propagating outside the MF. By adjusting the neighboring distance of the coils, a compact in-line polarizer or high-Q single-polarization resonator can be obtained. This stereo graphene-integrated resonator with unique 3D geometry eliminates the requirement of separate couplers, which makes the resonator especially attractive for special applications, such as gyroscopes and current sensors. This preliminary work could lay the foundation for future lab-on-a-rod GMF devices.

Figure 1 is a schematic of the GMF in-line polarizer (GMF-IP) or GMF single-polarization resonator (GMF-SR). The only difference between these lies in the distance (d, as labeled in Figure 1) between neighboring coils. To prevent the loss induced by the relatively high-index rod (polymethyl methacrylate or PMMA), a thin layer of low-index Teflon (Teflon® AF 601S1-100-6, DuPont, tens of micrometers in thickness) was initially dip-coated on the rod surface. A monolayer graphene sheet grown by chemical vapor deposition (CVD) was then mechanically transferred onto the surface of the Teflon coating. The length of the graphene sheet was carefully tailored so that it could be wrapped around the rod such that the opposite edges meet at the same line and do not overlap. Finally, an MF (~3 μm in diameter) was tightly wrapped on the graphene sheet supported by the rod.

The broken symmetry with respect to the local y = 0 plane causes the non-degeneracy of the two fundamental modes in this hybrid structure. Here, we define the mode with the dominant electric field component in the x-direction ($E_x$) as an even mode because $E_x$ is symmetric to the mirror plane (x = 0 plane, perpendicular to the local graphene plane). The other mode, mainly polarized in the y-direction, is considered an odd mode. Owing to the large evanescent field of the MF, a fraction of the light field penetrates the ambient environment, including the graphene sheet. For intrinsic or slightly doped graphene ($|\mu| < h\omega/4\pi$), inter-band transition dominates the high-frequency dynamic conductivity within the range from visible to infrared (IR).[7] Thus, the existence of the graphene sheet contributes to the losses in different modes, causing the device to exhibit a contrasting extinction ratio (ER).

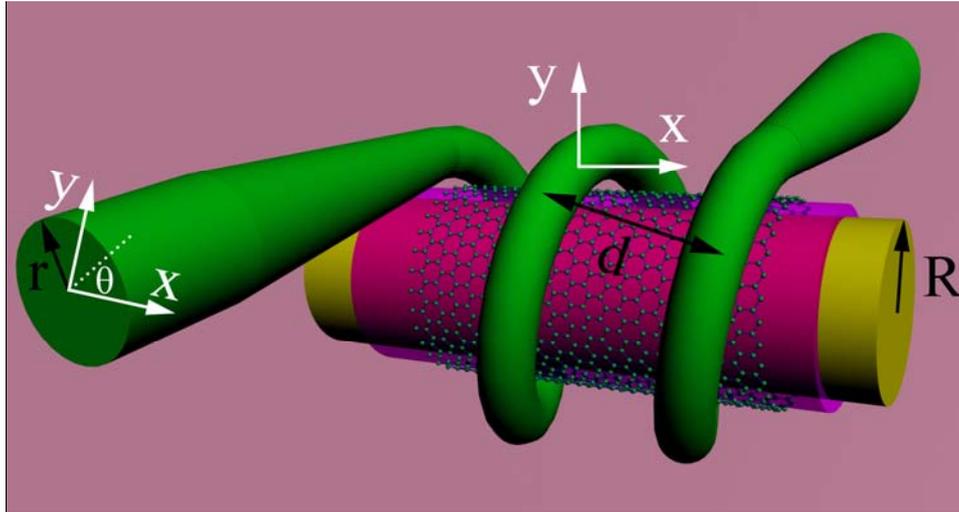

**Figure 1.** Schematic of a graphene-based MF in-line polarizer with a two-coil structure. The rod's diameter is hundreds of times larger than that of the MF. The GMF interaction length can be extended nearly indefinitely on a small piece of graphene by increasing the number of helical turns. R: radius of the rod; r: radius of the MF; $\theta$: angle between the incident polarization direction and x-axis. Two sets of local coordinates are shown, one at the input and the other in the middle.

We first studied the optical transmission properties of the GMF-IP, in which a distance (tens of micrometers) was maintained between different coils of the MF to prevent mutual coupling.[14] The extinction ratio was experimentally measured in the near infrared (NIR) range, from 1200 nm to 1650 nm. We chose this range because many high-order modes appear below 1200 nm, and the broadband light source (SuperK Versa, NKT) terminates near 1650 nm. By incorporating a linear polarizer followed by a half-wave plate between the source and the device, the polarization state of the incident light can be controlled by rotating the plate. As clearly shown in Figure 2, maximum transmission is achieved at $\theta = 90°$, in which case the odd mode is excited, whereas minimum transmission occurs at $\theta = 0°$. The device with one-coil structure (meaning that the MF is in contact with the graphene sheet over ~6.3 mm) gives an ER of ~5 dB at 1310 nm and ~8 dB at 1550 nm. As the operating wavelength increases, the ER increases; this phenomenon will be explained below. To further increase ER, a longer MF was employed to form a two-coil (Figure 3a) or multi-coil structure. As expected, ER nearly doubled when the contact between the MF and graphene was lengthened to ~12.6 mm. We note a dip in the output spectrum around 1380 nm, which is possibly caused by the absorption loss of the extra OH groups released during hydrogen flame heating.[15] A control experiment was also performed on an MF wrapped on a rod pretreated only with Teflon coating, without a graphene sheet. The results show no evident polarizing effect.

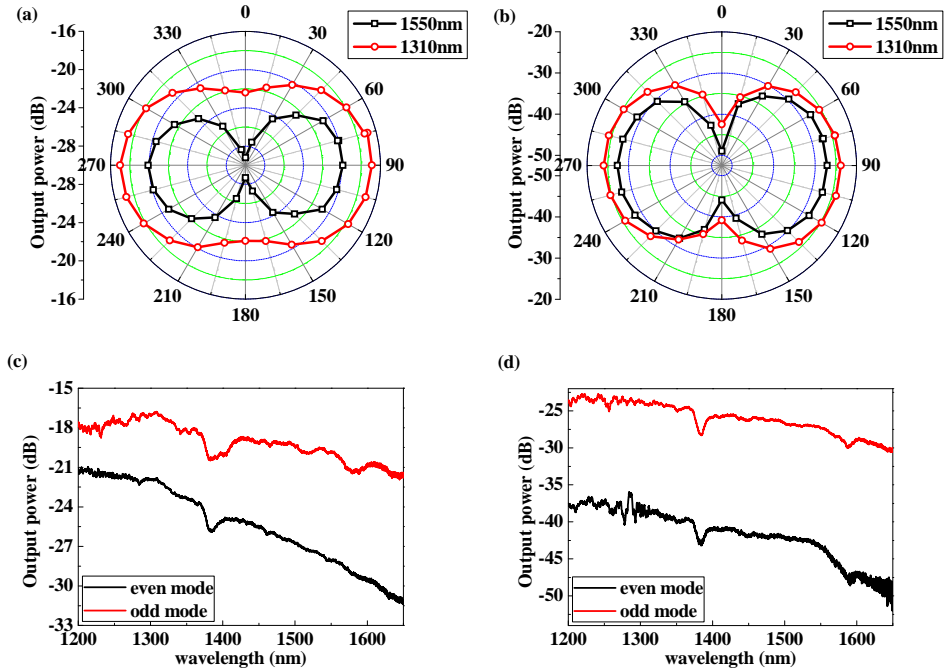

**Figure 2.** Output power as a function of $\theta$ at 1310 nm and 1550 nm for a (a) one-coil and (b) two-coil structure. The data was recorded by rotating the half-wave plate in 7.5° increments. Wavelength-dependent output power for a (c) one-coil and (d) two-coil structure. The black (red) line represents the data at $\theta = 0°$ (90°).

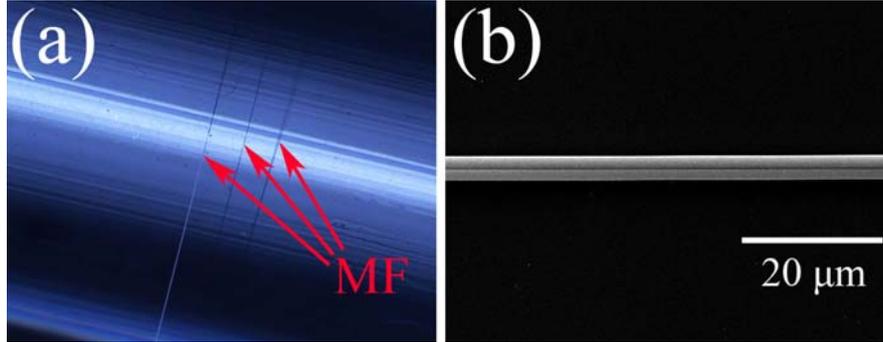

**Figure 3.** (a) Optical microscope image of a two-coil MF wrapped on a rod pretreated with Teflon coating and a graphene sheet. (b) SEM image of two coils close to each other.

When adjacent coils of the MF are sufficiently close, they form a high-Q resonator.[16] We carefully wrapped two MF coils onto the graphene-coated rod with the help of a microscope to ensure that the coils were close to each other, as illustrated in Figure 3b. The transmission spectra of GMF-SR for two orthogonal modes (even and odd) are shown in Figure 4. They clearly differ in output power because the modes suffer from different propagation losses. From Figure 4b, we find that around the telecommunication wavelength of 1550 nm, the free spectral range (FSR) and full width at half-maximum (FWHM) are approximately 0.23 nm and 0.08 nm, respectively, indicating that the Q-factor approaches $2\times10^4$. The difference in output power between the dips of the two modes is ~11 dB, from which we conclude that the device could work as an SR. In this manner, mutual coupling between different modes could be well suppressed, which is especially attractive for certain specialized applications such as gyroscopes and current sensors. This functionality cannot be realized in a one-dimensional fiber system. The extinction ratio could be further improved if the MFs are drawn into even thinner structures.

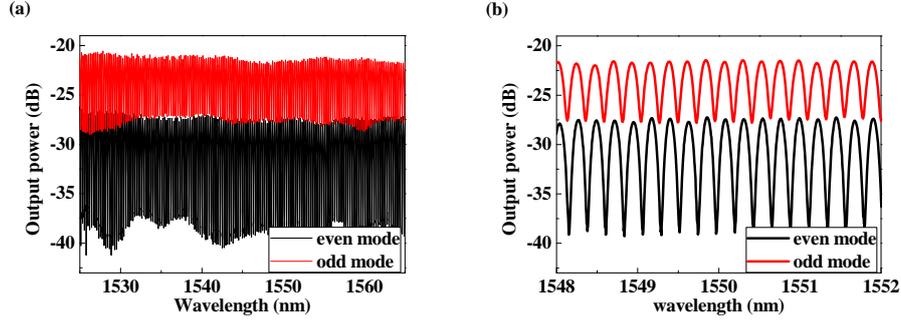

**Figure 4.** (a) Transmission spectrum of the graphene-based resonator for two orthogonal modes. (b) Detailed view around 1550 nm.

To confirm the experimental results, we theoretically analyzed the mode losses. In our simulations, the ultra-thin graphene sheet was treated as a d=1 nm-thick dielectric layer. The relative permittivity of graphene was obtained using the relation $\varepsilon = 1 + i\sigma/\varepsilon_0 \omega d$, where $\sigma$ is the dynamical conductivity determined from the Kubo formula, $\varepsilon_0$ is the permittivity of vacuum, and $\omega$ is the operating frequency.[7] The theoretical ER was obtained via $(\alpha_{even} - \alpha_{odd})L_{MF}$, with $\alpha_i$ (i = even or odd) being the attenuation constant of each mode and $L_{MF}$ being the length of the MF in contact with the graphene sheet. Figure 5a displays the results of ER versus operating wavelength for GMF-IP. The experimental results agree well with the theoretical predictions, except for some mismatch in the short wavelength range. This mismatch can be attributed to the contribution of high-order modes and the non-uniformity of MFs, both of which are neglected in our calculation. Moreover, ER increases with the wavelength, which is also observed in Bao et al.[7] Another calculation was performed to obtain the time-averaged power flow ratio in the graphene layer ($\eta_{graphene}$), shown in Figure 5b. For the even mode, we clearly see that as the wavelength increases, $\eta_{graphene}$ increases. For the odd mode, however, $\eta_{graphene}$ remains constant over the 450 nm wavelength range. Because the graphene sheet is a lossy medium, more energy will be attenuated when $\eta_{graphene}$ is high. This explains the increasing tendency of ER in Figure 5a.

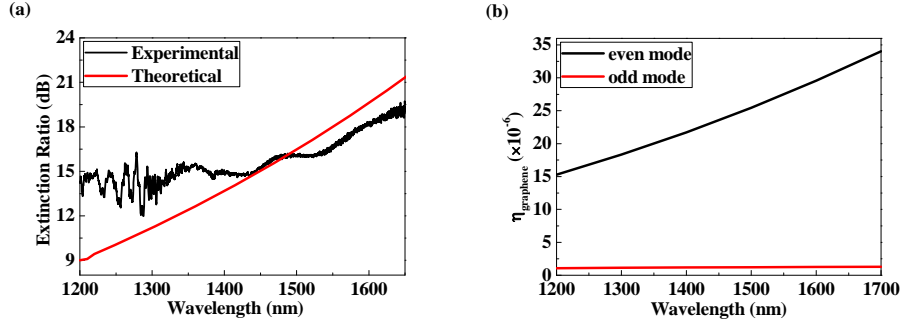

**Figure 5.** (a) Experimental and theoretical results of ER for GMF-IP. (b) Calculated time-averaged power flow ratio for power carried by the graphene sheet as a function of wavelength.

For GMF-SR, the transmission spectrum of each mode can be obtained through the following equation, which incorporates loss.[17]

$$T_i = \frac{e^{-4\pi R\beta_{2,i}} + \sin^2(2\pi R\kappa_i) - 2\sin(2\pi R\beta_{1,i})\sin(2\pi R\kappa_i)e^{-2\pi R\beta_{2,i}}}{e^{4\pi R\beta_{2,i}} + \sin^2(2\pi R\kappa_i) - 2\sin(2\pi R\beta_{1,i})\sin(2\pi R\kappa_i)e^{2\pi R\beta_{2,i}}},$$

where $\beta_{1,i}$ and $\beta_{2,i}$ are the real and imaginary parts of the complex propagation constant, respectively, and $\kappa_i$ is the coupling coefficient for the corresponding mode. Here, a semi-theoretical method was employed. $\beta_{1,i}$ and $\beta_{2,i}$ were calculated by the finite element method, while $2\pi R\kappa_{even} = 0.15$ and $2\pi R\kappa_{odd} = 1.0$ were chosen to fit the experimental results. These values of $\kappa_i$ were selected on the basis of two aspects. On one hand, $\kappa_i$ is dependent on polarization. On the other hand, $\kappa_i$ is difficult to predict because it is sensitive to both the distance between neighboring coils and the MF radius. In the fitting, the insertion loss ($I_i$) of different modes resulting from coupling, connection, scattering, and bending was not included. The calculated FSR was ~0.25 nm, which is close to the experimentally observed value (~0.23 nm). The even mode exhibits a lower transmission than the odd mode because in the even mode, the electric field, mainly polarized in the x-direction, interacts more with the graphene sheet.

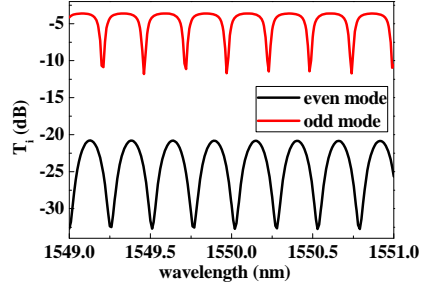

**Figure 6.** Calculated transmission of the GMF-SR based on a semi-theoretical method. Insertion loss resulting from coupling, connection, scattering, and bending was not taken into account here.

In conclusion, we have presented a new platform for miniaturizing GMF-integrated stereo-devices. The unique geometry enables practically unlimited light-graphene interaction lengths on a small piece of graphene, and the realization of a number of GMF broadband devices, including microresonators. As an example, we demonstrated the in-line manipulation of polarization with GMF-integrated devices. Firstly, a GMF-IP was shown to achieve an ER of 8 dB/coil. By employing a two-coil structure, an ER as high as ~16 dB was obtained over a 400 nm bandwidth in the telecommunication wavelength range. Second, we took a step toward the realization of a high-Q graphene-based SR with excellent suppression of polarization noise and used a semi-empirical model to explain the results. Similar results are expected with the use of RGB fibers in the visible range and endless single-mode photonic crystal fibers over the visible to NIR range. Moreover, our design also gives insights into the realization of future lab-on-a-rod devices.

Author information

Corresponding Author

*E-mail: feixu@nju.edu.cn

yqlu@nju.edu.cn

Author Contributions

The manuscript was written through contributions of all authors. All authors have given approval to the final version of the manuscript. ‡These authors contributed equally.

Funding Sources

This work is supported by National 973 program under contract No. 2012CB921803 and 2011CBA00205, National Science Fund for Excellent Young Scientists Fund (61322503) and National Science Fund for Distinguished Young Scholars (61225026). The authors also acknowledge the support from PAPD and the Fundamental Research Funds for the Central Universities.

Notes

The authors declare no competing financial interest.

REFERENCES

1. Bonaccorso, F.; Sun, Z.; Hasan, T.; Ferrari, A. C. *Nature Photon.* **2010,** 4, (9), 611-622.
2. Avouris, P. *Nano Lett.* **2010,** 10, (11), 4285-4294.
3. Meng, C.; Xiao, Y.; Wang, P.; Zhang, L.; Liu, Y.; Tong, L. *Advanced Materials* **2011,** 23, (33), 3770-3774.
4. Wo, J.; Wang, G.; Cui, Y.; Sun, Q.; Liang, R.; Shum, P. P.; Liu, D. *Opt. Lett.* **2012,** 37, (1), 67-69.
5. Xia, F.; Mueller, T.; Lin, Y.-m.; Valdes-Garcia, A.; Avouris, P. *Nat. Nanotech.* **2009,** 4, (12), 839-843.
6. Mueller, T.; Xia, F.; Avouris, P. *Nature Photon.* **2010,** 4, (5), 297-301.
7. Bao, Q.; Zhang, H.; Wang, B.; Ni, Z.; Lim, C. H. Y. X.; Wang, Y.; Tang, D. Y.; Loh, K. P. *Nat. Photon.* **2011,** 5, (7), 411-415.
8. Min, B.; Ostby, E.; Sorger, V.; Ulin-Avila, E.; Yang, L.; Zhang, X.; Vahala, K. *Nature* **2009,** 457, (7228), 455-458.
9. Sun, Z.; Hasan, T.; Torrisi, F.; Popa, D.; Privitera, G.; Wang, F.; Bonaccorso, F.; Basko, D. M.; Ferrari, A. C. *ACS Nano* **2010,** 4, (2), 803-810.
10. Koester, S. J.; Mo, L. *IEEE J. Sel. Top. Quant. Electron.* **2014,** 20, (1), 84-94.
11. Li, W.; Chen, B.; Meng, C.; Fang, W.; Xiao, Y.; Li, X.; Hu, Z.; Xu, Y.; Tong, L.; Wang, H.; Liu, W.; Bao, J.; Shen, Y. R. *Nano Letters* **2014,** 14, (2), 955-959.
12. Liu, Z.-B.; Feng, M.; Jiang, W.-S.; Xin, W.; Wang, P.; Sheng, Q.-W.; Liu, Y.-G.; Wang, D. N.; Zhou, W.-Y.; Tian, J.-G. *Laser Phys. Lett.* **2013,** 10, (6), 065901.
13. Kopp, V. I.; Churikov, V. M.; Singer, J.; Chao, N.; Neugroschl, D.; Genack, A. Z. *Science* **2004,** 305, (5680), 74-75.
14. Kou, J.-l.; Chen, Y.; Xu, F.; Lu, Y.-q. *Opt. Express* **2012,** 20, (27), 28431-28436.
15. Osanai, H.; Shioda, T.; Moriyama, T.; Araki, S.; Horiguchi, M.; Izawa, T.; Takata, H. *Electron. Lett.* **1976,** 12, (21), 549-550.
16. Brambilla, G. *J. Opt.* **2010,** 12, (4), 043001.
17. Sumetsky, M. *Opt. Express* **2004,** 12, (10), 2303-2316.